\documentclass[aps,pra,superscriptaddress,twocolumn,showpacs,floatfix]{revtex4}

\usepackage{amsmath}
\usepackage{graphicx}
\usepackage{bm}
\usepackage{mathrsfs}

\newcommand{\mno}{\mathnormal{\Omega}}
\newcommand{\mng}{\mathnormal{\Gamma}}

\begin{document}

\title{Nonlinear Propagation of Incoherent Photons in a Radiation Background}  

\author{Padma K.\ Shukla}
\affiliation{Institut f\"ur Theoretische Physik IV, Fakult\"at f\"ur
  Physik und Astronomie, Ruhr-Universit\"at Bochum, D--44780 Bochum,
  Germany}
\affiliation{Department of Physics, Ume{\aa} University, SE--901 87
  Ume{\aa}, Sweden}

\author{Mattias Marklund}
\affiliation{Department of Electromagnetics, Chalmers University of
  Technology, SE--412 96 G\"oteborg, Sweden}

\author{Gert Brodin}
\affiliation{Department of Physics, Ume{\aa} University, SE--901 87
  Ume{\aa}, Sweden}

\author{Lennart Stenflo}
\affiliation{Department of Physics, Ume{\aa} University, SE--901 87
  Ume{\aa}, Sweden}

\date{\today}

\begin{abstract}
The nonlinear propagation of intense incoherent photons in a
photon gas is considered. The photon-photon interactions
are governed by a pair of equations comprising
a wave-kinetic equation for the incoherent photons in
the presence of the slowly varying energy density
perturbations of sound-like waves, and an equation 
for the latter waves in a background where the  
photon coupling is caused by  
quantum electrodynamical effects. The coupled equations are 
used to derive a dispersion relation, which admits 
new classes of modulational instabilities of incoherent
photons. The present instabilities can lead to 
fragmentation of broadband short photon pulses in
astrophysical and laboratory settings.
\end{abstract}

\maketitle

The advent of quantum electrodynamics (QED) is one of the major
scientific achievements of the $20^{\mathrm{th}}$ century. It is an 
experimentally well confirmed theory, and it has
predicted a number of phenomena which were not expected in previous studies, 
e.g., the Casimir effect. Moreover, as opposed to the
classical theory of Maxwell, electromagnetic waves can according to QED
interact in the absence of a material mediator. 
Due to the possibility of exchanging virtual
electron--positron pairs, there is the intriguing effect of photon--photon
scattering, first discussed even before the interaction between light and
matter was well understood \cite{Heisenberg-Euler,Weisskopf}, 
and later derived within QED by Schwinger \cite{Schwinger}. The fact 
that strong electromagnetic fields can interact in vacuum, opens 
up for very interesting applications in extreme astrophysical
environments, such as magnetars \cite{magnetar}, where 
photon splitting and lensing may take place
\cite{Heyl-Hernquist1,Heyl-Hernquist2,Shaviv-Heyl-Lithwick,Denisov-Svertilov}.
The prospect of direct detection  
of the effect has also been discussed in the literature, and a number of 
suggestions for experimental setups have been given, involving second harmonic 
generation \cite{Ding-Kaplan2}, self-focusing \cite{Soljacic-Segev}, nonlinear
wave mixing in cavities
\cite{Brodin-Marklund-Stenflo1,Brodin-Marklund-Stenflo2} and  
waveguide propagation \cite{Brodin-etal}, respectively. Apart from the
astrophysical 
regions where high fields exist, the rapid development
of ultra-high laser fields \cite{Mourou-Barty-Perry,Pukhov} and related 
laser-plasma techniques \cite{Bulanov-Esirkepov-Tajima}, gives hope that 
the critical field strengths will be produced in laboratories. 

We shall below investigate the effect of incoherence of
high-frequency photons propagating on a radiation fluid
background. This will extend the work presented in
Ref. \cite{Marklund-Brodin-Stenflo}, and investigated numerically in
Ref. \cite{Shukla-Eliasson}, to the random phase regime. For this purpose,
a wave kinetic theory for high-frequency photons coupled to an acoustic wave
equation for a radiation fluid is presented. It is shown that
modulational instabilities are inherent in the system of
equations. Moreover, in the limit of the slow time-variation
approximation, we obtain a Vlasov equation with self-interaction 
for the high-frequency photons.

We consider an incoherent non-thermal high-frequency spectrum of
photons. As will be shown, this spectrum will be able to interact with
low-frequency acoustic-like perturbations. The high-frequency part is
treated by means of a wave kinetic description, whereas the
low-frequency part is 
described by an acoustic wave equation with a driver 
\cite{Marklund-Brodin-Stenflo} which follows from a radiation fluid
description.  
Let $N_k(t,\bm{r})$ denote the high frequency photon distribution function, 
normalised such that the corresponding number density is given by 
$n = \int\,N_k\,d^3k$. Then $N_k$ will satisfy the wave kinetic equation
\cite{Mendonca}
\begin{equation}
  \frac{\partial N_{k}}{\partial t} + 
  \bm{v}_g\cdot\frac{\partial N_{k}}{\partial\bm{r}} -
  \frac{\partial\omega_k}{\partial\bm{r}}%
  \cdot\frac{\partial N_{k}}{\partial\bm{k}} = 0
\label{eq:kinetic}
\end{equation}
where $\bm{v}_g = \partial\omega_k/\partial\bm{k}$ is the group
velocity, and 
\begin{equation}
  \omega_k = ck\left(1 -
  \tfrac{2}{3}\lambda\mathscr{E}\right) , 
\label{eq:dispersion-relation}
\end{equation}
where $\mathscr{E}$ is the radiation fluid density 
\cite{Bialynicka-Birula,Marklund-Brodin-Stenflo}, and
$\lambda=8\kappa$ or $14\kappa$, depending on the polarisation state
of the photon.  
Here $\kappa \equiv 2\alpha^2\hbar^3/45m_e^4c^5 \approx 1.63\times
10^{-30}\, \mathrm{m}\mathrm{s}^{2}/\mathrm{kg}$, $\alpha$ is
the fine-structure
constant, $2\pi \hbar$ the Planck constant, $m_e$ the electron mass, and
$c$ the velocity of light in vacuum. The dispersion relation
(\ref{eq:dispersion-relation}) is 
valid as long as there is no pair creation and the field strength is
smaller than the QED critical field, i.e., 
\begin{equation}
  \omega \ll m_ec^2/\hbar \,\,\text{ and }\,\, |\bm{E}| \ll
  E_{\text{crit}} \equiv 
  m_ec^2/e\lambda_c   
\end{equation}
respectively. Here $e$ is the elementary charge, $\lambda_c$ is the
Compton wave length, and $E_{\text{crit}} \simeq
10^{18}\,\mathrm{V}/\mathrm{m}$.

The high-frequency photons drive
low-frequency acoustic perturbations according to
\cite{Marklund-Brodin-Stenflo} 
\begin{equation}
  \left( \frac{\partial^2}{\partial t^2} - \frac{c^2}{3}\nabla^2
  \right)\mathscr{E} =
  -\frac{2\lambda\mathscr{E}_0}{3}\left(
  \frac{\partial^2}{\partial t^2} + c^2\nabla^2 \right) \int \hbar\omega_k
  N_k\, d^3k 
\label{eq:wave}
\end{equation}
where the constant $\mathscr{E}_0$ is the \emph{background} radiation
fluid energy density. This hybrid description, where the high-frequency 
part is treated kinetically, and the low-frequency part is described within 
a fluid theory, applies when the mean-free path between 
photon-photon collisions is shorter than the wavelengths of the low-frequency 
perturbations.  We note that the intensity $I_k = \hbar\omega_k
N_k/\epsilon_0$ satisfies Eq.\ (\ref{eq:kinetic}), and is normalised
such that $\langle|E|^2\rangle = \int\,I_k\,d^3k$, where $E$ is the high-
frequency electric field strength, and $\epsilon_0$ is the dielectric constant of vacuum.  
The equations presented here 
resemble the photon--electron system in
Ref. \cite{Shukla-Stenflo-PoP}, where the interaction between random phase
photons and sound waves in an electron--positron plasma has been
investigated.   

Next we consider a small low-frequency long wavelength perturbation of
a homogeneous  
background spectrum, i.e. $N_k = N_{k0} + N_{k1}\exp [i(Kz -
  {\mno}t)]$, $N_{k1} 
\ll N_{k0}$, and $ \mathscr{E} = \mathscr{E}_1\exp[i(Kz - {\mno}t)]$ and 
linearise our equations. We thus obtain (using expression
(\ref{eq:dispersion-relation}) for $\omega_k$, and introducing
$\hat{\bm{k}}={\bm{k}}/k$) 
\begin{subequations}
\begin{equation}
  N_{k1} = \frac{2\lambda k\mathscr{E}_1}{3}\frac{Kc}{{\mno} -
  Kc\hat{\bm{k}}\cdot\hat{\bm{z}}}\hat{\bm{z}}\cdot\frac{\partial
  N_{k0}}{\partial\bm{k}} 
\end{equation}
and
\begin{equation}
  \mathscr{E}_1 = -\frac{2\lambda c\hbar\mathscr{E}_0}{3}\frac{{\mno}^2 +
  K^2c^2}{{\mno}^2 - K^2c^2/3}\int k N_{k1}\,d^3k ,
\end{equation}
\end{subequations}
which, when combined, give the nonlinear dispersion relation
\begin{equation}
  1 = -\frac{\mu K}{3}\frac{{\mno}^2 +
  K^2c^2}{{\mno}^2 - K^2c^2/3}\int\frac{k^2}{{\mno}
  -Kc\hat{\bm{k}}\cdot\hat{\bm{z}}}\hat{\bm{z}}\cdot\frac{\partial
  N_{k0}}{\partial\bm{k}} \,d^3k ,
\label{eq:kinetic-dispersion}
\end{equation}
where we have introduced the constant $\mu =
\frac{4}{3}\lambda^2c^2\hbar\mathscr{E}_0$. 
As was found in Ref. \cite{Shukla-Stenflo-PoP,Shukla-Stenflo-PLA}, we may
have growth for a large class of background distributions $N_{k0}$. 

\begin{figure}[t]
\begin{center}
  \includegraphics[width=.49\textwidth]{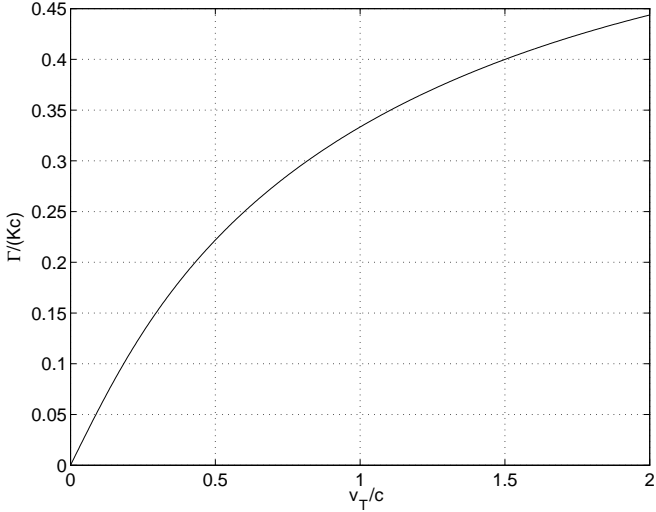}
\end{center}
\caption{The transverse instability for the mono-energetic
  case. $\mng/Kc$ plotted as a function of 
  $v_T/c$, as given in Eq.\ (\ref{eq:growth1}).} 
\label{fig1}
\end{figure}

(a) For a mono-energetic high
frequency background, we have $N_{k0} = n_0\delta(\bm{k} - \bm{k}_0)$.
The nonlinear dispersion relation (\ref{eq:kinetic-dispersion})
then reduces to 
\begin{eqnarray}
  && ({{\mno}^2 - K^2c^2/3}){({\mno} -
  Kc\cos\theta_0)^2} = \tfrac{1}{3}\mu n_0k_0 K\times
\nonumber \\ &&\quad 
 \times({{\mno}^2 +
  K^2c^2})
  [{Kc + (2{\mno} -3Kc\cos\theta_0)\cos\theta_0}] ,
\label{eq:disprel-mono}
\end{eqnarray} 
where we have introduced $\cos\theta_0 \equiv
\hat{\bm{k_0}}\cdot\hat{\bm{z}}$. This mono-energetic background has a
transverse instability when $\theta_0 = \pi/2$, with the
growth rate 
\begin{equation}
  {\mng} =  \tfrac{1}{\sqrt{6}}Kc\left[ \sqrt{ \left(\frac{v_T}{c}\right)^4
  + 14\left(\frac{v_T}{c}\right)^2 + 1} - \left(\frac{v_T}{c}\right)^2 - 1
  \right]^{1/2} ,
\label{eq:growth1}
\end{equation}
where ${\mng} \equiv -i{\mno}$, and $v_T^2 \equiv
\mu n_0k_0c$, and $v_T$ is a characteristic speed of the
system. The expression in the 
square bracket is positive definite. In Fig.\ \ref{fig1} we have
plotted ${\mng}/Kc$ as a function of $v_T/c$.

\begin{figure}[t]
\begin{center}
  \includegraphics[width=.49\textwidth]{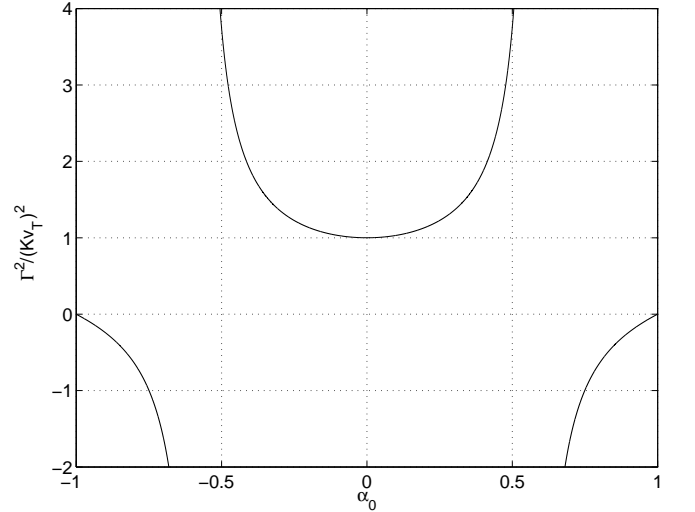}
\end{center}
\caption{$(\mng/Kv_T)^2$, according to Eq.\ (\ref{eq:growth2}),
  plotted as a function of
  $\alpha_0 = \cos\theta_0$ in the mono-energetic case.}
\label{fig2}
\end{figure}

In fact, under most circumstances, $v_T \ll c$. Using the 
expression (\ref{eq:disprel-mono}), we then have two
branches. The branch corresponding to $\mno \simeq
Kc/\sqrt{3}$ is always stable for small $v_T$, while for the branch
corresponding to $\mno \simeq Kc\cos\theta_0$ we obtain the
growth rate  
\begin{equation}
  \mng = Kv_T\sqrt{\frac{1 - \cos\theta_0}{1 - 3\cos\theta_0}} ,
\label{eq:growth2}
\end{equation} 
which is consistent with (\ref{eq:growth1}) in the limit
$\theta_0 \rightarrow \pi/2$. In Fig.\
\ref{fig2}, the behaviour of the growth rate 
(\ref{eq:growth2}) is depicted. 

(b) The high-frequency photons have generally a 
spread in momentum space. For simplicity we 
here choose the background intensity distribution as a
shifted Gaussian, i.e. 
\begin{equation}
  I_{k0} = \frac{\mathscr{I}_0}{\pi^{3/2}k_W^3}\exp\left[
  - \frac{(\bm{k} - \bm{k}_0)^2}{k_W^2}\right] ,
\end{equation}
where $\mathscr{I}_0 = \langle|E_0|^2\rangle$ 
is the (constant) background intensity and $k_W$
is the width of the distribution around $\bm{k}_0$. Then the
dispersion relation is 
\begin{eqnarray}
   1 &=& -\frac{b^2}{k_W^5}  
  \frac{\eta^2 + 1}{\eta^2 - 1/3} 
  \int\bigg[\frac{k(k_0\cos\theta_0 - k\cos\theta)}{\eta
   - \cos\theta}
\nonumber \\ &&\qquad
\times\exp\left( -\frac{k^2 -
  2\bm{k}\cdot\bm{k}_0}{k_W^2} \right) \bigg]d^3k ,
\label{eq:disprel-gaussian}
\end{eqnarray}
where
$b^2 = (4/9\pi^{3/2})\lambda^2\epsilon_0\mathscr{E}_0
\mathscr{I}_0\exp(-k_0^2/k_W^2)$ and $\eta \equiv {\mno}/Kc$. 


Assuming that the deviation of $\bm{k}_0$ from the $\hat{\bm{z}}$-axis
is small, and that $\delta \equiv k_0/k_W \ll 1$, we can integrate Eq.\
(\ref{eq:disprel-gaussian}), keeping terms linear in $\delta$, to
obtain   
\begin{eqnarray}
  1 &\simeq& -\pi b^2\frac{\eta^2 + 1}{\eta^2 - 1/3}\Bigg[ 
  \frac{3\sqrt{\pi}}{2} + 8\delta\eta\cos\theta_0
\nonumber \\ && \!\!\!\!\!\!\!\!\!\!\!\!\!\!\!\! \!\!\!\!\!\!\!\!
  +
  \left(\delta\cos\theta_0 - \frac{3\sqrt{\pi}}{4}\eta
  -4\delta\eta\cos\theta_0 \right)(2\,\text{arctanh}\,\eta - i\pi)  
  \Bigg] ,
\end{eqnarray}
for $0 < \eta < 1$. Thus, we see that the non-zero
width of the distribution complicates the characteristic behaviour of
the dispersion relation by a considerable amount. It is clear though,
that the width will introduce a reduction of the growth rate, as
compared to the mono-energetic case. 


We may also look at the case when the time-dependence is weak, i.e.
$\partial^2\mathscr{E}/\partial t^2 \ll c^2\nabla^2\mathscr{E}$,
such that Eq.\ (\ref{eq:wave}) yields 
\begin{equation}
  \mathscr{E} = 2\lambda\mathscr{E}_0\int\hbar\omega_kN_k\,d^3k .
\label{eq:energy}
\end{equation}
Upon using Eq.\ (\ref{eq:energy}) in (\ref{eq:dispersion-relation}), we find
\begin{equation}
  \frac{\partial\omega_k}{\partial\bm{r}} =
  -\mu k
  \frac{\partial}{\partial\bm{r}}\int\,k'N_{k'}\,d^3k' 
\end{equation}
Hence Eq.\ (\ref{eq:kinetic}) becomes
\begin{equation}
  \frac{\partial N_{k}}{\partial t} + 
  \bm{v}_g\cdot\frac{\partial N_{k}}{\partial\bm{r}} +
  \mu k
  \left(\frac{\partial}{\partial\bm{r}}\int k'N_{k'}\,d^3k'\right)%
  \cdot\frac{\partial N_{k}}{\partial\bm{k}} = 0 ,
\end{equation}
which in the one-dimensional case reduces to
\begin{equation}
  \frac{\partial N_{k}}{\partial t} + v_g\frac{\partial
  N_{k}}{\partial x} + \mu k\left(\frac{\partial}{\partial x}\int
  k'N_{k'}\,dk'\right)\frac{\partial N_{k}}{\partial k} = 0 .
\label{eq:self}
\end{equation}
A similar equation may of course be derived for the intensity
$I_k$. 

Equation (\ref{eq:self}) gives the evolution of high-frequency photons
on a slowly varying background radiation fluid, and it may be used to
analyse the long term behaviour of amplitude modulated intense short
incoherent laser pulses.

We have considered the nonlinear propagation of
randomly distributed intense short photon pulses in a photon
gas. The photon-photon interactions are described by means of
a QED model, in which an ensemble of incoherent photon pulses
is governed by a wave kinetic equation where the coupling between
the intense photon pulses and the sound-like waves of the photon gas 
is due to slow variations of the sound wave energy 
distribution.  The intense photon pressure, in turn, modifies the sound 
wave propagation. The wave kinetic and the sound wave equations 
form a closed system, which has been used to derive a dispersion 
relation. By choosing appropriate spectra for short pulse photons, we
analyze the dispersion relation to show the existence 
of new classes of modulational instabilities. The
latter can cause fragmentation of incoherent
photon pulses in astrophysical contexts and in forthcoming experiments using
very intense short laser pulses.



\begin{thebibliography}{99}

\bibitem{Heisenberg-Euler} W.\ Heisenberg and H.\ Euler, Z.\
  Phys.~\textbf{98} 714 (1936).

\bibitem{Weisskopf} V.S.\ Weisskopf, K.\ Dan.\
  Vidensk.\ Selsk.\ Mat.\ Fy.\ Medd.~\textbf{14} 1 (1936).

\bibitem{Schwinger} J.\ Schwinger, Phys.\ Rev.~\textbf{82}
  664 (1951).

\bibitem{magnetar} C.\ Kouveliotou \textit{et al.}, Nature
  \textbf{393} 235 (1998).

\bibitem{Heyl-Hernquist1} J.S.\ Heyl and L.\ Hernquist, J.\ Phys.\ A:
  Math.\ Gen.\ \textbf{30} 6485 (1997).
 
\bibitem{Heyl-Hernquist2} J.S.\ Heyl and L.\ Hernquist, Phys.\ Rev.\ D
  \textbf{55} 2449 (1997). 

\bibitem{Shaviv-Heyl-Lithwick} N.J.\ Shaviv, J.S.\ Heyl and Y.\
  Lithwick, MNRAS \textbf{306} 333 (1999).

\bibitem{Denisov-Svertilov} V.I.\ Denisov and S.I.\ Svertilov,
  Astron.\ Astroph.\ \textbf{399} L39 (2003).

\bibitem{Ding-Kaplan2}
  Y.J.\ Ding and A.E.\ Kaplan,
  Phys.\ Rev.\ Lett.~\textbf{63} 2725 (1989).

\bibitem{Soljacic-Segev} M.\ Solja\v{c}i\'c and M.\ Segev,
  Phys.\ Rev.\ A \textbf{62} 043817 (2000).

\bibitem{Brodin-Marklund-Stenflo1} G.\ Brodin, M.\ Marklund and L.\
  Stenflo, Phys.\ Rev.\ Lett.~\textbf{87} 171801 (2001).

\bibitem{Brodin-Marklund-Stenflo2}
  G.\ Brodin, M.\ Marklund and L.\
  Stenflo, Phys.\ Scripta \textbf{T98} 127 (2002).

\bibitem{Brodin-etal} 
  G.\ Brodin, L.\ Stenflo, D.\ Anderson, M.\
  Lisak, M.\ Marklund and P.\ Johannisson, Phys.\ Lett. A
  \textbf{306} 206 (2003).

\bibitem{Mourou-Barty-Perry} G.A.\ Mourou, C.P.J.\ Barty and M.D.\
  Perry, Phys.\ Today \textbf{51} 22 (1998).
 
\bibitem{Pukhov} A.\ Pukhov, Rep.\ Prog.\ Phys.\ \textbf{66} 47
  (2003).

\bibitem{Bulanov-Esirkepov-Tajima} S.V.\ Bulanov, T.\ Esirkepov and
  T.\ Tajima, Phys.\ Rev.\ Lett.\ \textbf{91} 085001 (2003).

\bibitem{Marklund-Brodin-Stenflo} M.\ Marklund, G.\ Brodin and L.\
  Stenflo, Phys.\ Rev.\ Lett.\ \textbf{91}  163601 (2003).

\bibitem{Shukla-Eliasson} P.K.\ Shukla and B.\ Eliasson, Modulational
  and filamentational instabilities of intense photon pulses and their
  dynamics in a photon gas, submitted.

\bibitem{Mendonca} J.T.\ Mendon\c{c}a, \textit{Theory of Photon
  Acceleration} (Institute of Physics Publishing, Bristol, 2001).

\bibitem{Bialynicka-Birula} Z.\ Bialynicka--Birula and I.\
  Bialynicki--Birula, Phys.\ Rev.\ D \textbf{2} 2341 (1970).

\bibitem{Shukla-Stenflo-PoP} P.K.\ Shukla and L.\ Stenflo, Phys.\ Plasmas
  \textbf{5} 1554 (1998).

\bibitem{Shukla-Stenflo-PLA} P.K.\ Shukla and L.\ Stenflo, Phys.\
  Lett.\ A \textbf{237} 385 (1998).

\end{thebibliography}
\end{document}